\documentclass[final,11pt]{article}

\usepackage{multirow}
\usepackage[nottoc]{tocbibind}
\usepackage{geometry}
\usepackage{subcaption}
\usepackage[affil-it, auth-sc]{authblk}
\usepackage[english]{babel}
\usepackage[utf8]{inputenc}
\usepackage{cite}
\usepackage{graphicx}
\usepackage{amssymb}
\usepackage{amsthm}
\usepackage{amsmath}
\usepackage{amsfonts}
\usepackage{verbatim}
\usepackage{lineno}
\usepackage{todonotes}
\presetkeys{todonotes}{inline, color=blue!30, size=\small}{}

\usepackage{color, soul}
\definecolor{darkred}{rgb}{0.9, 0.0, 0.0}
\definecolor{darkgreen}{rgb}{0.0, 0.5, 0.0}

\usepackage[colorlinks,bookmarks,linkcolor=darkred, citecolor=darkgreen]{hyperref}
\usepackage{eso-pic}
\usepackage[sort&compress,numbers]{natbib}
\usepackage{doi}
\usepackage{paralist,epsfig}
\usepackage{relsize}
\usepackage{nicefrac,esint}
\usepackage{float}
\usepackage{cleveref}

\def\slash#1{#1\!\!\!/\!\,\,}

\begin{document}

\AddToShipoutPictureFG*{
    \AtPageUpperLeft{\put(-60,-60){\makebox[\paperwidth][r]{LA-UR-21-30638,~FERMILAB-PUB-21-739-T}}}  
    }
    
\title{Radiative (anti)neutrino energy spectra from muon, pion, and kaon decays}
\author{Oleksandr Tomalak \thanks{tomalak@lanl.gov}}

\affil{Theoretical Division, Los Alamos National Laboratory, Los Alamos, NM 87545, USA \vspace{1.2mm}}
\affil{Theoretical Physics Department, Fermi National Accelerator Laboratory, Batavia, IL 60510, USA \vspace{1.2mm}}
\affil{Department of Physics and Astronomy, University of Kentucky, Lexington, KY 40506, USA \vspace{1.2mm}}

\date{\today}

\maketitle

To describe low-energy (anti)neutrino fluxes in modern coherent elastic neutrino-nucleus scattering experiments as well as high-energy fluxes in precision-frontier projects such as the Enhanced NeUtrino BEams from kaon Tagging (ENUBET) and the Neutrinos from STORed Muons (nuSTORM), we evaluate (anti)neutrino energy spectra from radiative muon ($\mu^- \to e^- \bar{\nu}_e \nu_\mu (\gamma),~\mu^+ \to e^+ {\nu}_e \bar{\nu}_\mu (\gamma)$), pion $\pi_{\ell 2}$ ($\pi^- \to \mu^- \bar{\nu}_\mu (\gamma),~\pi^+ \to \mu^+ {\nu}_\mu (\gamma)$), and kaon $K_{\ell 2}$ ($K^- \to \mu^- \bar{\nu}_\mu (\gamma),~K^+ \to \mu^+ {\nu}_\mu (\gamma)$) decays. We compare detailed $\mathrm{O} \left( \alpha \right)$ distributions to the well-known tree-level results, investigate electron-mass corrections and provide energy spectra in analytical form. Radiative corrections introduce continuous and divergent spectral components near the endpoint, on top of the monochromatic tree-level meson-decay spectra, which can change the flux-averaged cross section at $6\times 10^{-5}$ level for the scattering on $^{40}\mathrm{Ar}$ nucleus with (anti)neutrinos from the pion decay at rest. Radiative effects modify the expected (anti)neutrino fluxes from the muon decay around the peak region by $3-4$ permille, which is a precision goal for next-generation artificial neutrino sources.

\tableofcontents

\section{Introduction}
\label{sec1}

Muon, pion, and kaon decays provide us with basic information on the low-energy effective field theories of the Standard Model and serve as main sources for the corresponding low-energy constants. Thus, precise lifetime measurements give us access to the Fermi coupling constant $\mathrm{G}_\mathrm{F}$, the pion decay constant $f_\pi$, and the product of the kaon decay constant $f_K$ with Cabibbo-Kobayashi-Maskawa (CKM) matrix element $V_{us}$. These decays show an excellent agreement between modern experiments with precise Standard Model predictions~\cite{Ott:1971rs,Numao:1995qf,Koptev:1995je,KLOE:2007wlh,FAST:2007rsc,MuLan:2010shf,Casella:2013bla}. 

Muon decay is a pure leptonic process independent of hadron-structure corrections. At leading order, the muon decays to electron and two corresponding (anti)neutrinos. Radiative decay is suppressed by the electromagnetic coupling constant and is a sub-leading process for a sizable photon energy resolution. The next subdominant channel also contains an $e^+e^-$ pair, in addition to the electron and neutrinos. Its branching is suppressed by 5 orders of magnitude compared to the main decay mechanism (the branching ratio is $\left(3.4 \pm 0.4 \right) \times 10^{-5}$~\cite{SINDRUM:1985vbg,ParticleDataGroup:2020ssz}). Other decay channels of the muon were not observed experimentally. 

The dominant pion decay mode contains a muon and a muon (anti)neutrino in the final state. The subdominant electronic channel is helicity suppressed by 4 orders of magnitude~\cite{PiENu:2015seu}. The corresponding radiative decays and decays accompanied with $e^+e^-$ pairs are suppressed by the electromagnetic coupling constant. The other observed channel $\pi^+ \to e^+ \nu_e \pi^0$ is suppressed by a small phase-space volume by a factor $\left(1.036\pm0.006\right)\times10^{-8}$~\cite{Pocanic:2003pf,ParticleDataGroup:2020ssz} at the level of decay width. 

Having a larger mass, kaons have a variety of decay modes. Pure hadronic, pure leptonic, and semi-leptonic modes are allowed. The dominant leptonic decay with the branching ratio $\left(63.56\pm0.11\right)~\%$ has a muon and a corresponding (anti)neutrino in the final state. Decays into the neutral pion are not suppressed by the phase-space volume compared to decays of $\pi^\pm$ and have branchings around 5~\% for $K_{e3}$ mode $K^\pm \to \pi^0 e^\pm \nu_e$ and around 3.4~\% for $K_{\mu3}$ mode $K^\pm \to \pi^0 \mu^\pm \nu_\mu$. Other decay channels with electrons are suppressed by five orders of magnitude.

Muon, pion, and kaon decay channels are at the heart of artificial (anti)neutrino sources. At low energies, experiments dedicated to searches for sterile neutrinos~\cite{LSND:2001aii,CAPTAIN:2013irr,CAPTAIN:2020pup,CCM:2021leg,Athar:2021xsd} and to measurements of the coherent elastic neutrino-nucleus scattering cross sections~\cite{COHERENT:2017ipa,COHERENT:2019iyj,COHERENT:2020iec,Barbeau:2021exu} are performed with (anti)neutrinos from stopped muons and pions~\cite{LSND:1996jxj,Maschuw:1998qh,Conrad:2009mh,Lazauskas:2010rh,Alonso:2010fs,LSND:2001aii,Pagliaroli:2012hq,Brice:2013fwa,CAPTAIN:2013irr,COHERENT:2017ipa,Baxter:2019mcx}. At higher energies, neutrino oscillation experiments~\cite{T2K:2011qtm,Hyper-KamiokandeProto-:2015xww,T2K:2019bcf,NOvA:2019cyt,DUNE:2020ypp} are based on muons, pions, and kaons decaying in flight~\cite{Kopp:2006ky,T2K:2012bge,Adamson:2015dkw,MINERvA:2016iqn}. Achieving percent-level precision at low and high energies requires an improved control over the (anti)neutrino production mechanisms. In particular, radiative corrections to muon and meson decays can distort (anti)neutrino energy spectra~\cite{Tomalak:2020zfh}.

While uncertainties on the electron antineutrino fluxes could be reduced from 10~\% to a few-\% level relying on the well-known inverse beta decay reaction at low energies~\cite{Vogel:1999zy,Kurylov:2002vj,Fukugita:2004cq,Raha:2011aa,Ankowski:2019tbc,COHERENT:2021xhx}, the muon component has no feasible ``standard candles". However, the precise theoretical knowledge of muon and pion decays could allow us to control the muon over electron flavor ratios of incoming (anti)neutrinos. While radiative corrections in muon, pion, and kaon decays are a relatively old subject, only corrections to the total decay width and to distributions w.r.t. charged lepton and photon variables are well-described in the literature~\cite{Behrends:1955mb,Kinoshita:1957zz,Berman:1958ti,Kinoshita:1958ru,Bailin:1964vky,Matsson:1969ai,Grotch:1968zza,Roos:1971mj,Girotti:1971gb,Ross:1973ry,Mohammad:1976qd,Greub:1993kg,Czarnecki:1994pu,Czarnecki:1994bn,Konyshev:1995cg,Bauberger:1997zz,Ferroglia:1999tg,vanRitbergen:1998yd,vanRitbergen:1999fi,Davydychev:2000ee,Arbuzov:2001ui,Fischer:2002hn,Davydychev:2002ni,Arbuzov:2002pp,Arbuzov:2002rp,Arbuzov:2004wr,Anastasiou:2005pn,Pak:2008qt,Caola:2014daa,Fael:2020tow,Czakon:2021ybq,Berman:1958gx,Kinoshita:1959ha,Ginsberg:1963zz,Kamal:1964zz,Bailin:1970ie,Goldman:1976gh,Marciano:1976jc,Snellman:1968dpw,Garcia:1981cy,Bryman:1982et,Jezabek:1988ja,Holstein:1990ua,Gabrielli:1992np,Bijnens:1992en,Greub:1993kg,Marciano:1993sh,Finkemeier:1994ev,Czarnecki:1994pu,Finkemeier:1995gi,Neufeld:1995mu,Kuraev:1996vn,Chen:1997gf,Chen:2008mg,Chen:2010ue,Cirigliano:2001mk,Broncano:2002hs,Descotes-Genon:2005wrq,Cirigliano:2007xi,Cirigliano:2007ga,Cirigliano:2008wn,Sirlin:2012mh}. To quantify the (anti)neutrino fluxes, another set of observables with (anti)neutrino kinematics are of interest. In this work, we evaluate leading in the electromagnetic coupling constant $\alpha$ corrections ($\mathrm{O} \left( \alpha \right)$) to the (anti)neutrino energy spectra and compare our calculations with previous results for the muon decay~\cite{Greub:1993kg,Broncano:2002hs} in the approximation of vanishing electron mass as well as with well-known results for the total decay widths~\cite{Kinoshita:1958ru,Nir:1989rm,Bijnens:1992en,vanRitbergen:1999fi,Pak:2008qt}. Our work paves the way for control of the muon over electron flavor ratios for incoming (anti)neutrino fluxes at the percent level and below.

The need for improved neutrino-nucleus cross-section inputs for modern and future neutrino oscillation experiments~\cite{Jen:2014aja,Benhar:2015wva,Ankowski:2016bji,Ankowski:2016jdd,Katori:2016yel,NuSTEC:2017hzk,Nagu:2019uco,Rocco:2020jlx,CLAS:2021neh} has motivated new cross-section measurements with potentially known fluxes. For example, the Enhanced NeUtrino BEams from kaon Tagging (ENUBET) project aims to control the (anti)neutrino fluxes by tagging charged leptons from decays of light mesons~\cite{Longhin:2014yta,Meregaglia:2016vxf}, kaons and pions in this case. Reconstruction of (anti)neutrino fluxes from high-statistics charged lepton measurements could potentially reduce flux uncertainties from $5$-$10$~\% level, when estimates are based on the hadroproduction cross sections, to a percent level or even below. Alternatively, (anti)neutrino fluxes can be precisely controlled via decays of large samples of stored muons, when only two types of (anti)neutrinos are present as decay products. Likewise, the absolute flux normalization can be controlled by precise measurements of the stored currents. This technique is the basis for the Neutrinos from STORed Muons (nuSTORM) activity~\cite{nuSTORM:2012jbd,nuSTORM:2013cqr,Adey:2013afh}, which will potentially allow percent-level cross-section measurements. The target precision goals call for the improved theoretical description of muon and light-meson decays.

The paper is organized as follows. In Section~\ref{sec2}, we describe the decay of charged pions and kaons. We provide details of tree-level energy spectra as well as virtual and real contributions at leading order in chiral expansion. We quantify uncertainties of resulting spectra, present relative contributions to (anti)neutrino energy distributions from radiative corrections, and estimate the size of strong dynamics effects. In Section~\ref{sec3}, we calculate radiative corrections to (anti)neutrino energy spectra from muon decay, study the relative size of $\mathrm{O} \left(\alpha\right)$ contributions compared to the leading-order result, and present electron-mass corrections. We provide conclusions and outlook in Section~\ref{sec4}.

\section{Pion and kaon decays}
\label{sec2}

Leptonic decays of charged light mesons result in the well-established source of monochromatic (anti)neutrinos when the (anti)neutrino energy depends only on masses of the light meson and charged lepton. The radiation of photons adds smooth component to (anti)neutrino energy distributions. In this Section, we calculate (anti)neutrino energy spectra from decays of light mesons in the meson rest frame, accounting for possible radiation of one photon and estimating the size of structure-dependent contributions. We consider pion $\pi_{\ell 2}:~\pi^- \to \mu^- \bar{\nu}_\mu \left(\gamma\right)$ and kaon $K_{\ell 2}:~K^- \to \mu^- \bar{\nu}_\mu \left(\gamma\right)$ decays. Charge-conjugated reactions $\pi^+ \to \mu^+ {\nu}_\mu (\gamma)$ and $K^+ \to \mu^+ {\nu}_\mu (\gamma)$ are obtained by replacing antineutrinos with neutrinos in all following discussions, while electron-flavor decays are given by replacing the muon mass with the electron mass. We derive all results by considering the example of the pion decay. For numerical evaluations, we substitute the pion mass and pion low-energy constants with the kaon mass and corresponding kaon constants.

\subsection{Pion and kaon decays at tree level}
\label{sec2:pi_tree}

At the level of quarks, the pion decay $\pi^- \left( p_\pi \right) \to \mu^- \left( p_\mu \right) \bar{\nu}_\mu \left( k_{\bar{\nu}_\mu}\right)$ is described by the low-energy four-fermion Lagrangian
\begin{align}
    {\cal L}_{\rm eff} = -  c^{u d} \bar{\mu} \gamma_\nu \mathrm{P}_\mathrm{L} \nu_{\mu}
  \, \bar{{u}} \gamma^\nu \mathrm{P}_\mathrm{L}  d + \mathrm{h.c.},
\end{align}
where the effective coupling constant is expressed in terms of the Fermi coupling $\mathrm{G}_\mathrm{F}$ and the CKM matrix element $V_{u d}$ as $c^{u d} = 2 \sqrt{2} \mathrm{G}_\mathrm{F} V_{u d} + \mathrm{O} \left( \frac{\alpha}{\pi} \right)$, c.f.~\cite{Descotes-Genon:2005wrq,Hill:2019xqk} for a detailed determination at order $\mathrm{O} \left( \alpha \right)$. For low-energy applications, mesons are the appropriate degrees of freedom with the corresponding leading-order Lagrangian
\begin{align}
    {\cal L}_{\rm eff} =
    -   \sqrt{2} \mathrm{G}_\mathrm{F} V_{ud} f_\pi 
  \bar{\mu} \gamma_\nu \mathrm{P}_\mathrm{L} \nu_{\mu} \mathrm{D}^\nu \pi^- + \mathrm{h.c.}, \label{eq:pion_decay_lagrangian1}
\end{align}
with the covariant derivative $\mathrm{D}^\nu = \partial^\nu + i e A^\nu$, where $e^2 = 4 \pi \alpha$, $A^\nu$ is the photon field, and the pion-decay constant $f_\pi$ is defined from the matrix element of the quark current
\begin{align}
c^{u d} <0 | \bar{{u}} \gamma^\nu \mathrm{P}_\mathrm{L}  d  | \pi^-\left( p_\pi \right)> = - i \sqrt{2} \mathrm{G}_\mathrm{F} V_{ud} f_\pi p_\pi^\nu.
\end{align}
Exploiting equations of motion, the leading-order Lagrangian is expressed as a derivative-free expression
\begin{align}
    {\cal L}_{\rm eff} =
   i \sqrt{2} \mathrm{G}_\mathrm{F} V_{ud} f_\pi m_\mu
  \bar{\mu} \mathrm{P}_\mathrm{L} \nu_{\mu}\pi^- + \mathrm{h.c.}. \label{eq:pion_decay_lagrangian2}
\end{align}
The squared matrix element at leading order $|\mathrm{T}_\mathrm{LO}|^2 =  4 \mathrm{G}_\mathrm{F}^2 |V_{ud}|^2 f^2_\pi m^2_\mu  k_{\bar{\nu}_\mu} \cdot p_\mu$ results in the following expression for the decay width $\Gamma_\mathrm{LO}$ in the pion rest frame with the monochromatic (anti)neutrino spectrum:
\begin{align}
\Gamma_\mathrm{LO} \left( \pi^- \to \mu^- \bar{\nu}_\mu \right) = \frac{\mathrm{G}_\mathrm{F}^2 |V_{ud}|^2 f^2_\pi}{8 \pi} m^2_\mu m_\pi \left( 1- \frac{m^2_\mu}{m^2_\pi} \right)^2 \int \delta \left( E_{\bar{\nu}_\mu} - \frac{m_\pi^2-m_\mu^2}{2 m_\pi} \right) \mathrm{d} E_{\bar{\nu}_\mu}. \label{eq:tree_level_pion_decay_width}
\end{align}
This equation represents the well-known helicity suppression of the electron decay channel compared to the dominant $\pi_{\mu 2}$ decay.

At leading order, the radiative pion decay $\pi^- \left( p_\pi \right) \to \mu^- \left( p_\mu \right) \bar{\nu}_\mu \left( k_{\bar{\nu}_\mu}\right) \gamma \left( k_\gamma \right)$ is described by the structure-independent gauge-invariant inner Bremsstrahlung contribution $\mathrm{T}^{1\gamma}_\mathrm{IB}$~\cite{Berman:1958gx,Kinoshita:1959ha,Ginsberg:1963zz,Bryman:1982et}:
\begin{align}
\mathrm{T}^{1\gamma}_\mathrm{IB} =  - e \sqrt{2} \mathrm{G}_\mathrm{F} V_{ud} f_\pi m_\mu \bar{\mu} \left( \frac{ \gamma^\nu \slash{k}_\gamma}{ 2  p_\mu \cdot k_\gamma } + \frac{p^\nu_\mu}{ p_\mu \cdot k_\gamma  }- \frac{p_\pi^\nu}{ p_\pi \cdot k_\gamma }\right) \mathrm{P}_\mathrm{L} \nu_{\mu} \pi^- \varepsilon^\star_\nu,
\end{align}
where $\slash{k} \equiv k_\mu \gamma^\mu$ for any four-vector $k$. At next order of the chiral expansion $\mathrm{O} \left( \frac{m^2_\pi}{16 \pi^2 f^2_\pi}\right)$, strong dynamics contribute to the radiation of the real photon with a matrix element $\mathrm{T}^{1\gamma}_\mathrm{SD}$~\cite{Bijnens:1992en}:
\begin{align}
\mathrm{T}^{1\gamma}_\mathrm{SD} = - i e \frac{\sqrt{2} \mathrm{G}_\mathrm{F} V_{ud}}{m_\pi} \left( F_V \varepsilon^{\nu \lambda \alpha \beta} \left( k_\gamma \right)_\alpha \left( p_\pi \right)_\beta + i F_A \left( k_\gamma \cdot p_\pi g^{\lambda \nu} - k_\gamma^\lambda p_\pi^\nu \right)\right) \bar{\mu} \gamma_\lambda \mathrm{P}_\mathrm{L} \nu_{\mu}  \varepsilon^\star_\nu, \label{eq:structure_dependent}
\end{align}
with nonperturbative vector $F_V$ and axial $F_A$ form factors. Form factors are functions of the squared momentum transfer $q^2 = \left( p_\pi - k_\gamma \right)^2$. Up to $\mathrm{O} \left( \frac{m^2_\pi}{16 \pi^2 f^2_\pi}\right)$, these form factors do not depend on the kinematics. Leading contributions in the chiral expansion are expressed in terms of Gasser and Leutwyler low-energy constants~\cite{Gasser:1983yg,Gasser:1984gg} as~\cite{Bijnens:1992en}
\begin{align}
F_V &= \frac{m_\pi}{4 \pi^2 f_\pi} + \mathrm{O} \left( \frac{m^2_\pi}{16 \pi^2 f^2_\pi}\right), \\
F_A &= \frac{8 m_\pi}{f_\pi} \left( L_9 + L_{10} \right) + \mathrm{O} \left( \frac{m^2_\pi}{16 \pi^2 f^2_\pi}\right),
\end{align}
with the natural size of low-energy constants $L_9,~L_{10} \sim \frac{1}{\left( 4 \pi \right)^2}$. The resulting radiation is determined by the sum of these two contributions $ \mathrm{T}^{1\gamma} = \mathrm{T}^{1\gamma}_\mathrm{IB} + \mathrm{T}^{1\gamma}_\mathrm{SD}$.

In this work, we calculate the (anti)neutrino energy spectra from the inner Bremsstrahlung $\mathrm{T}^{1\gamma}_\mathrm{IB}$ and consider the structure-dependent contributions $\mathrm{T}^{1\gamma}_\mathrm{SD}$ from $F_V$ and $F_A$ as an alternative error estimate vs $\mathrm{O} \left( \frac{m^2_\pi}{16 \pi^2 f^2_\pi}\right)$ Chiral perturbation theory (ChPT) power counting. Note that the complete inclusion of radiative effects at $\mathrm{O} \left( \frac{m^2_\pi}{16 \pi^2 f^2_\pi}\right)$ will require introducing an additional structure-dependent form factor for the calculation of virtual corrections to Eq.~(\ref{eq:structure_dependent}) that will be $\mathrm{O} \left( \frac{m^2_\pi}{16 \pi^2 f^2_\pi}\right)$ uncertain due to poorly-known low-energy constants for electroweak pion loops at $\mathrm{O} \left( \frac{m^2_\pi}{16 \pi^2 f^2_\pi}\right)$~\cite{Cirigliano:2007ga}.

\subsection{Virtual corrections}
\label{sec2:pi_virtual}

At the order of interest $\mathrm{O} \left( \frac{m^2_\pi}{16 \pi^2 f^2_\pi} \right)$, virtual radiative corrections enter as a factor $f$ multiplying the tree-level amplitude $\mathrm{T}_\mathrm{LO}$, i.e., virtual corrections change the leading-order result as
\begin{align}
\mathrm{T}_\mathrm{LO} \to \left( 1 + \frac{\alpha}{\pi} f + \mathrm{O} \left( \frac{\alpha^2}{\pi^2},~ \frac{m^2_\pi}{16 \pi^2 f^2_\pi}  \frac{\alpha}{\pi}\right) \right) \mathrm{T}_\mathrm{LO}.
\end{align}
Starting with the Lagrangian of Eq.~(\ref{eq:pion_decay_lagrangian1}), the factor $f$ is determined by field renormalization constants for external charged particles $Z_\pi,~Z_\mu$ and diagrams with virtual photon exchange between pion and muon lines as well as with an exchange between the contact 5-point interaction and charged particle lines $f^v$:
\begin{align}
f = f^v + \frac{\pi}{\alpha} \left( \sqrt{Z_\pi Z_\mu} -1 \right).
\end{align}
Starting with the Lagrangian of Eq.~(\ref{eq:pion_decay_lagrangian2}), the same correction $f^v$ is obtained from the diagram with virtual photon exchange between pion and muon lines and QED counterterm for the muon mass~\cite{Kinoshita:2016drm}.

The field renormalization factors are evaluated from the one-loop self energies in the $\overline{\mathrm{MS}}$ renormalization scheme as~\cite{Vanderhaeghen:2000ws,Heller:2018ypa,Heller:2019dyv,Tomalak:2019ibg}
\begin{align}
Z_\mu &= 1 -  \frac{\alpha}{4\pi} \frac{\xi_\gamma}{\varepsilon} - \frac{\alpha}{4\pi} \left(  \ln \frac{\mu^2}{m_\mu^2} + 2\ln \frac{\lambda^2}{m_\mu^2} + 4 \right) + \frac{\alpha}{4\pi} \left( 1 - \xi_\gamma  \right) \left( \ln \frac{\mu^2}{\lambda^2} + 1 + \frac{a \xi_\gamma \ln  a \xi_\gamma }{1 - a \xi_\gamma }  \right),  \label{eq:charged_lepton_Z_factor} \\
Z_\pi &= 1  -  \frac{\alpha}{4\pi} \frac{1+\xi_\gamma}{\varepsilon} - \frac{\alpha}{4\pi} \left(   \ln \frac{\mu^4}{m^4_\pi} + 2 \ln \frac{\lambda^2}{m^2_\pi}+ 4 \right)+ \frac{\alpha}{4\pi} \left( 1 - \xi_\gamma  \right) \left( \ln \frac{\mu^2}{\lambda^2} + 1 + \frac{a \xi_\gamma \ln  a \xi_\gamma }{1 - a \xi_\gamma }  \right), 
\end{align}
with the renormalization scale in dimensional regularization $\mu$, where the number of dimensions is $D=4-2 \varepsilon$, the photon mass $\lambda$ regulates the infrared divergence, $\xi_\gamma$ is the photon gauge-fixing parameter, and $a$ is an arbitrary constant, which enters the photon propagator in the momentum space as
\begin{align}
\mathrm{\Pi}^{\mu \nu} \left( k_\gamma \right) = \frac{i}{k^2_\gamma - \lambda^2} \left( - g^{\mu \nu} + \left( 1 - \xi_\gamma \right) \frac{k_\gamma^{\mu} k_\gamma^{\nu}}{k^2_\gamma - a \xi_\gamma \lambda^2 }\right). \label{eq:photon_propagator}
\end{align}
Adding the virtual contribution $f^v$, the dependence on parameters $\xi_\gamma$ and $a$ cancels. The resulting correction is expressed in terms of the pion and muon masses as
\begin{align}
 f = -\frac{7}{4} \ln \frac{\mu}{m_\mu} - \frac{3}{2} +  \left( 1+ \frac{m_\pi^2 + m_\mu^2}{m_\pi^2 - m_\mu^2} \ln \frac{m_\mu}{m_\pi} \right)\ln \frac{m_\pi}{\lambda} - \frac{1}{2} \frac{m_\pi^2 + m_\mu^2}{m_\pi^2 - m_\mu^2} \ln \frac{m_\mu}{m_\pi} \left( 1 - \ln \frac{m_\mu}{m_\pi}\right).
\end{align}
The virtual correction depends on the renormalization scale $\mu$. Therefore, the low-energy constant in Eqs.~(\ref{eq:pion_decay_lagrangian1}) and (\ref{eq:pion_decay_lagrangian2}) also depends on the scale. To avoid new definitions or fixing the renormalization scale, we integrate over the total phase-space of radiated photons and express all our results in terms of the experimental decay width of the pion 
\begin{align}
\Gamma^{\mathrm{exp}}_{\pi^- \to \mu^- \bar{\nu}_\mu \left( \gamma \right)} = \Gamma \left( \pi^- \to \mu^- \bar{\nu}_\mu \right) + \Gamma \left( \pi^- \to \mu^- \bar{\nu}_\mu \gamma \right).
\end{align}

\subsection{Soft-photon Bremsstrahlung}
\label{sec2:pi_soft}

Radiative decay with photons of arbitrary small energy cannot be experimentally distinguished from the decay without the radiation. Consequently, all events with photons below some energy cutoff $k_\gamma \le \varepsilon$ (in the pion rest frame) contribute to the measured observables. This region of the phase space also cancels the infrared-divergent contributions from virtual diagrams. The corresponding radiative decay width factorizes in terms of the tree-level decay width of Eq.~(\ref{eq:tree_level_pion_decay_width}) as
\begin{align}
  \Gamma \left( \pi^- \to \mu^-  \bar{\nu}_\mu  \gamma \left( k_\gamma \le \varepsilon \right) \right) =  \frac{\alpha}{\pi} \delta_s \left( \varepsilon \right) \Gamma_\mathrm{LO} \left( \pi^- \to \mu^-  \bar{\nu}_\mu \right), \label{eq:soft_pion_decay_width}
\end{align}
with the correction $\delta_s \left( \varepsilon \right) $~\cite{Lee:1964jq,Aoki:1980ix,Sarantakos:1982bp,Passera:2000ug,Tomalak:2019ibg}, which is universal for all QED processes:
\begin{align}
\delta_s \left( \varepsilon \right) =  \frac{1}{ \beta}\left(  \mathrm{Li}_2 \frac{1-\beta}{1+\beta} - \frac{\pi^2}{6} \right)-  \frac{2}{\beta} \left( \beta -   \frac{1}{2} \ln \frac{1+\beta}{1-\beta} \right)\ln  \frac{2 \varepsilon}{\lambda}  + \frac{1}{2 \beta}  \ln \frac{1+\beta}{1-\beta}\left( 1  +  \ln \frac{\rho \left(1+ \beta \right) }{4 \beta^2}   \right)+1\,, \label{eq:soft_result}
\end{align}
where $\beta$ is the velocity of the muon in the pion rest frame: $\beta = \frac{m^2_\pi - m^2_\mu}{m^2_\pi + m^2_\mu}$ and $\rho=\sqrt{1-\beta^2} = \frac{2 m_\pi m_\mu}{m^2_\pi + m^2_\mu}$. As a result, soft and virtual contributions multiply the monochromatic tree-level spectrum of Eq.~(\ref{eq:tree_level_pion_decay_width}) with infrared-finite factor, i.e., independent of the fictitious photon mass $\lambda$~\cite{Bloch:1937pw,Nakanishi:1958ur,Kinoshita:1962ur,Lee:1964is}, as
\begin{align}
 \frac{\Gamma \left( \pi^- \to \mu^- \bar{\nu}_\mu \right) + \Gamma \left( \pi^- \to \mu^- \bar{\nu}_\mu \gamma \left( k_\gamma \le \varepsilon \right) \right)}{\Gamma_\mathrm{LO} \left( \pi^- \to \mu^- \bar{\nu}_\mu \right)} =   1 + \frac{\alpha}{\pi} \left( \delta_s \left( \varepsilon \right) + 2 f \right) . \label{eq:tree_level_soft_and_virtual_pion_decay_width}
\end{align}

\subsection{Contribution of hard photons}
\label{sec2:pi_hard}

The squared matrix elements for the radiative pion decay amplitudes $\mathrm{T}^{1\gamma}_\mathrm{IB}$ and $\mathrm{T}^{1\gamma}_\mathrm{SD}$ introduced in Section~\ref{sec2:pi_tree} are expressed in terms of Lorentz invariants as
\begin{align}
\frac{|\mathrm{T}^{1\gamma}_\mathrm{IB}|^2 }{e^2 |\mathrm{T}_\mathrm{LO}|^2} k_{\bar{\nu}_\mu} \cdot p_\mu &= - \left( \frac{p_\pi}{p_\pi \cdot k_\gamma} - \frac{p_\mu}{p_\mu \cdot k_\gamma} \right)^2 + \frac{k_{\bar{\nu}_\mu} \cdot p_\mu}{p_\mu \cdot k_\gamma} - \frac{k_{\bar{\nu}_\mu} \cdot p_\pi}{p_\pi \cdot k_\gamma} - m^2_\mu \frac{k_{\bar{\nu}_\mu} \cdot k_\gamma}{\left( p_\mu \cdot k_\gamma \right)^2} + \frac{k_{\bar{\nu}_\mu} \cdot k_\gamma}{p_\mu \cdot k_\gamma}\nonumber \\
&+ \frac{k_{\bar{\nu}_\mu} \cdot k_\gamma p_\pi \cdot p_\mu }{p_\mu \cdot k_\gamma p_\pi \cdot k_\gamma} , \label{eq:leading_radiation} \\
\frac{\left(\mathrm{T}^{1\gamma}_\mathrm{IB}\right)^\star \mathrm{T}^{1\gamma}_\mathrm{SD} + \left(\mathrm{T}^{1\gamma}_\mathrm{SD}\right)^\star \mathrm{T}^{1\gamma}_\mathrm{IB}}{e^2 |\mathrm{T}_\mathrm{LO}|^2} &= \frac{2 F_A}{f_\pi m_\pi}  \left( \frac{m^2_\pi k_{\bar{\nu}_\mu} \cdot k_\gamma}{p_\pi \cdot k_\gamma k_{\bar{\nu}_\mu} \cdot p_\mu} - \frac{k_{\bar{\nu}_\mu} \cdot p_\pi}{k_{\bar{\nu}_\mu} \cdot p_\mu} - \frac{ k_{\bar{\nu}_\mu} \cdot k_\gamma p_\pi \cdot p_\mu }{p_\mu \cdot k_\gamma k_{\bar{\nu}_\mu} \cdot p_\mu} + \frac{ p_\pi \cdot k_\gamma}{p_\mu \cdot k_\gamma} \right)\nonumber \\
&+ \frac{2 \left( F_A - F_V \right)}{f_\pi m_\pi} \frac{k_{\bar{\nu}_\mu} \cdot k_\gamma p_\pi \cdot k_\gamma}{p_\mu \cdot k_\gamma k_{\bar{\nu}_\mu} \cdot p_\mu}  , \\
\frac{|\mathrm{T}^{1\gamma}_\mathrm{SD}|^2}{e^2 |\mathrm{T}_\mathrm{LO}|^2} k_{\bar{\nu}_\mu} \cdot p_\mu &=  \frac{2\left( F_A - F_V \right)^2}{f_\pi^2 m_\pi^2 m^2_\mu}  k_{\bar{\nu}_\mu} \cdot k_\gamma  p_\pi \cdot k_\gamma p_\mu \cdot p_\pi +  \frac{2\left( F_A + F_V \right)^2}{f_\pi^2 m_\pi^2 m^2_\mu} p_\mu \cdot k_\gamma p_\pi \cdot k_\gamma  k_{\bar{\nu}_\mu} \cdot p_\pi \nonumber \\
&-  \frac{2 \left( F_A^2 + F_V^2 \right)}{f_\pi^2 m^2_\mu} k_{\bar{\nu}_\mu} \cdot k_\gamma  p_\mu \cdot k_\gamma.
\end{align}
Performing the full phase-space integration from $|\mathrm{T}^{1\gamma}_\mathrm{IB}|^2$, we relate the exclusive decay width of Eq.~(\ref{eq:tree_level_soft_and_virtual_pion_decay_width}) to the experimental decay width of the pion $\Gamma^{\mathrm{exp}}_{\pi^- \to \mu^- \bar{\nu}_\mu \left( \gamma \right)}$, up to contributions of order $\mathrm{O} \left( \frac{\alpha^2}{\pi^2},~\frac{m^2_\pi}{16\pi^2f^2_\pi}\frac{\alpha}{\pi}\right)$:
\begin{align}
 \Gamma \left( \pi^- \to \mu^- \bar{\nu}_\mu \right) + \Gamma \left( \pi^- \to \mu^- \bar{\nu}_\mu \gamma \left( k_\gamma \le \varepsilon \right) \right) =  \left( 1 - \frac{\alpha}{\pi} \delta \left( \varepsilon \right) \right) \Gamma^{\mathrm{exp}}_{\pi^- \to \mu^- \bar{\nu}_\mu \left( \gamma \right)},  \label{eq:tree_level_soft_and_virtual_pion_decay_width_in_terms_of_the_experiment}
 \end{align}
 with the scale-independent correction $\delta \left( \varepsilon \right)$~\cite{Kinoshita:1959ha}:
 \begin{align}
 \delta \left( \varepsilon \right) &= \left( 1+ \frac{m_\pi^2 + m_\mu^2}{m_\pi^2 - m_\mu^2} \ln \frac{m_\mu}{m_\pi} \right) \ln \frac{4\varepsilon^2}{m^2_\pi} - 2 \ln \left( 1 - \frac{m^2_\mu}{m^2_\pi} \right) + \frac{11}{4} + \left( \frac{9}{8} - \frac{3}{8} \frac{\left( m_\pi^2 + m_\mu^2 \right)^2}{\left( m_\pi^2 - m_\mu^2 \right)^2} \right) \ln \frac{m_\mu}{m_\pi} \nonumber \\
 &+ \frac{m_\pi^2 + m_\mu^2}{m_\pi^2 - m_\mu^2} \left( \mathrm{Li}_2  \frac{m^2_\mu}{m^2_\pi} - \frac{\pi^2}{6}  + \frac{3}{8} \left( \ln \frac{m_\mu^2}{m^2_\pi} -  1 \right)\right).
\end{align}
Note that the relative radiative correction to the total inclusive decay rate $\delta + \delta_s + 2 f$ has lepton-mass singularity
 \begin{align}
 \delta + \delta_s + 2 f \underset{m_\mu \ll m_\pi}{\longrightarrow}  3 \frac{\alpha}{\pi} \ln \frac{m_\mu}{m_\pi},
\end{align}
in agreement with~\cite{Berman:1958gx,Kinoshita:1959ha,Holstein:1990ua,Knecht:1999ag,Cirigliano:2007ga}. This singularity is multiplied by the lepton mass from the tree-level amplitude and is absent on the level of physical observables according to the Kinoshita-Lee-Naunberg theorem~\cite{Bloch:1937pw,Yennie:1961ad,Kinoshita:1962ur,Lee:1964is}.

All double-differential distributions can be derived by an appropriate change of variables from the double-differential energy spectrum
\begin{align}
\frac{\mathrm{d}  \Gamma \left( \pi^- \to \mu^- \bar{\nu}_\mu \gamma \right) }{\Gamma^{\mathrm{exp}}_{\pi^- \to \mu^- \bar{\nu}_\mu \left( \gamma \right)} \mathrm{d} E_{\bar{\nu}_\mu} \mathrm{d} E_\gamma} = \frac{\alpha}{\pi} \frac{m_\pi}{2 \omega^2_{\bar{\nu}_\mu} } \left(\frac{ 2 \omega_{\bar{\nu}_\mu}}{\omega_{\bar{\nu}_\mu} - E_{\bar{\nu}_\mu}} \frac{1 - \frac{E_{\bar{\nu}_\mu}}{m_\pi}}{E_\gamma} - \frac{\frac{m^2_\mu}{m^2_\pi}E_{\bar{\nu}_\mu}}{\left( \omega_{\bar{\nu}_\mu} - E_{\bar{\nu}_\mu} \right)^2}  -  \frac{ 1 - \frac{E_\gamma}{m_\pi} }{\omega_{\bar{\nu}_\mu} - E_{\bar{\nu}_\mu}}  -\frac{\omega_{\bar{\nu}_\mu}}{E_\gamma^2} - \frac{1}{m_\pi} \right), \label{eq:double_differential}
\end{align}
with the maximum value of the (anti)neutrino energy allowed by the kinematics $\omega_{\bar{\nu}_\mu} = \frac{m^2_\pi - m^2_\mu}{2 m_\pi}$, corresponding to the process without radiation. The double-differential distribution in Eq.~(\ref{eq:double_differential}) is in agreement with~\cite{Bijnens:1992en,Cirigliano:2007ga}. The leading continuous (anti)neutrino spectrum from the pion decay is expressed as
\begin{align}
\frac{\mathrm{d} \Gamma \left( \pi^- \to \mu^- \bar{\nu}_\mu \gamma \right) }{\Gamma^{\mathrm{exp}}_{\pi^- \to \mu^- \bar{\nu}_\mu \left( \gamma \right)} \mathrm{d} E_{\bar{\nu}_\mu}} &= \frac{\alpha}{\pi}  \frac{1}{\omega^2_{\bar{\nu}_\mu} }  \left( \frac{1}{2} \frac{E^2_{\bar{\nu}_\mu} + \frac{m^2_\mu}{4}}{m_\pi - 2 E_{\bar{\nu}_\mu}} - 2 E_{\bar{\nu}_\mu} -\frac{m^2_\mu}{8 m_\pi} - \frac{E_{\bar{\nu}_\mu} m^2_\mu}{4\left( m_\pi - 2 E_{\bar{\nu}_\mu} \right)^2} \right) \nonumber \\
&+ \frac{\alpha}{\pi}  \frac{1}{\omega^2_{\bar{\nu}_\mu} }   \frac{ 2 E^2_{\bar{\nu}_\mu} + \left( m_\pi - E_{\bar{\nu}_\mu} \right) \omega_{\bar{\nu}_\mu} \ln \left( 1 - \frac{2 E_{\bar{\nu}_\mu}}{m_\pi} \right) }{\omega_{\bar{\nu}_\mu} - E_{\bar{\nu}_\mu}}+ \mathrm{O} \left( \frac{\alpha^2}{\pi^2},~\frac{m^2_\pi}{16 \pi^2 f^2_\pi} \frac{\alpha}{\pi}  \right),
\end{align}
with the range of the (anti)neutrino energy $ 0 \le E_{\bar{\nu}_\mu} \le \omega_{\bar{\nu}_\mu} - \varepsilon$. Uncertainties due to the structure-dependent contributions are estimated from ChPT power counting by a factor of order $\mathrm{O} \left( \frac{m^2_\pi}{16 \pi^2 f^2_\pi}\right)$, as well as contributions from $F_V$ and $F_A$, the leading linear in form factors $\Gamma_{F} \left( \pi^- \to \mu^- \bar{\nu}_\mu \gamma \right)$ and subleading quadratic in form factors $\Gamma_{F^2} \left( \pi^- \to \mu^- \bar{\nu}_\mu \gamma \right)$ terms:
\begin{align}
\frac{\mathrm{d} \Gamma_F \left( \pi^- \to \mu^- \bar{\nu}_\mu \gamma \right) }{\Gamma^{\mathrm{exp}}_{\pi^- \to \mu^- \bar{\nu}_\mu \left( \gamma \right)} \mathrm{d} E_{\bar{\nu}_\mu}} &= \frac{\alpha}{\pi} \frac{ \omega_{\bar{\nu}_\mu} - E_{\bar{\nu}_\mu}}{\omega^2_{\bar{\nu}_\mu} }  \frac{m_\pi }{f_\pi}  \left( \frac{m_\pi - E_{\bar{\nu}_\mu}}{m_\pi - 2 E_{\bar{\nu}_\mu}} \frac{2 E_{\bar{\nu}_\mu}}{m_\pi}+ \ln  \left( 1 - \frac{2 E_{\bar{\nu}_\mu}}{m_\pi} \right) \right) F_A  \nonumber \\
&+ \frac{\alpha}{\pi} \frac{2}{3 f_\pi}  \frac{E_{\bar{\nu}_\mu}^2}{\omega^2_{\bar{\nu}_\mu}} \frac{\left( \omega_{\bar{\nu}_\mu} - E_{\bar{\nu}_\mu} \right)^2 }{\left(m_\pi - 2 E_{\bar{\nu}_\mu} \right)^2} \left( 1 + \frac{2m_\pi}{m_\pi - 2 E_{\bar{\nu}_\mu}} \right) \left( F_A - F_V \right) , \\
\frac{\mathrm{d} \Gamma_{F^2} \left( \pi^- \to \mu^- \bar{\nu}_\mu \gamma \right) }{\Gamma^{\mathrm{exp}}_{\pi^- \to \mu^- \bar{\nu}_\mu \left( \gamma \right)} \mathrm{d} E_{\bar{\nu}_\mu}} &= \frac{\alpha}{\pi} \frac{2 m_\pi }{f_\pi^2 m^2_\mu}  \frac{E_{\bar{\nu}_\mu}^2}{\omega^2_{\bar{\nu}_\mu}} \frac{\left( \omega_{\bar{\nu}_\mu} - E_{\bar{\nu}_\mu} \right)^3 }{\left(m_\pi - 2 E_{\bar{\nu}_\mu} \right)^2} \left( \left(m_\pi - E_{\bar{\nu}_\mu} \right) \left( F_A + F_V \right)^2 - m_\pi \left( F^2_A + F^2_V \right) \right) \nonumber \\
&+ \left( \left( E_{\bar{\nu}_\mu} -  \frac{m^2_\mu}{2m_\pi} \right) \frac{E^2_{\bar{\nu}_\mu}}{3} + \left(  m_\pi E_{\bar{\nu}_\mu} - \frac{m^2_\mu}{4} \right) \left(  \frac{m_\pi }{2} -  E_{\bar{\nu}_\mu} \right)+ \frac{m^2_\mu m^2_\pi / 8}{m_\pi - 2 E_{\bar{\nu}_\mu}}\right)    \nonumber \\
&\times \frac{\alpha}{\pi} \frac{2 m_\pi }{f_\pi^2 m^2_\mu}  \frac{E_{\bar{\nu}_\mu}}{\omega^2_{\bar{\nu}_\mu}} \frac{\left( \omega_{\bar{\nu}_\mu} - E_{\bar{\nu}_\mu} \right)^3 }{\left(m_\pi - 2 E_{\bar{\nu}_\mu} \right)^3}\left( F_A - F_V \right)^2.
\end{align}

\subsection{Decay spectra and integrated cross sections}
\label{sec2:pi_results}

In the following Fig.~\ref{fig:pion_kaon_decay}, we compare the leading-order continuous component of the (anti)neutrino energy spectrum $\mathrm{d} \Gamma \left( \pi^- \to \mu^- \bar{\nu}_\mu \gamma \right)/\mathrm{d} E_{\bar{\nu}_\mu}$ to structure-dependent power-suppressed contributions from the form factors $F_V$ and $F_A$: $\mathrm{d} \Gamma_{F} \left( \pi^- \to \mu^- \bar{\nu}_\mu \gamma \right)/\mathrm{d} E_{\bar{\nu}_\mu}$, $\mathrm{d} \Gamma_{F^2} \left( \pi^- \to \mu^- \bar{\nu}_\mu \gamma \right)/\mathrm{d} E_{\bar{\nu}_\mu}$ as well as to the power-counting error estimate. We take the ChPT low-energy constants $L_9$ and $L_{10}$ from Refs.~\cite{Geng:2003mt,Bijnens:2014lea,Unterdorfer:2008zz} and all  other numerical values for physical quantities from the Particle Data Group (PDG)~\cite{ParticleDataGroup:2020ssz}.
\begin{figure}[t]
          \centering
          \includegraphics[height=0.289\textwidth]{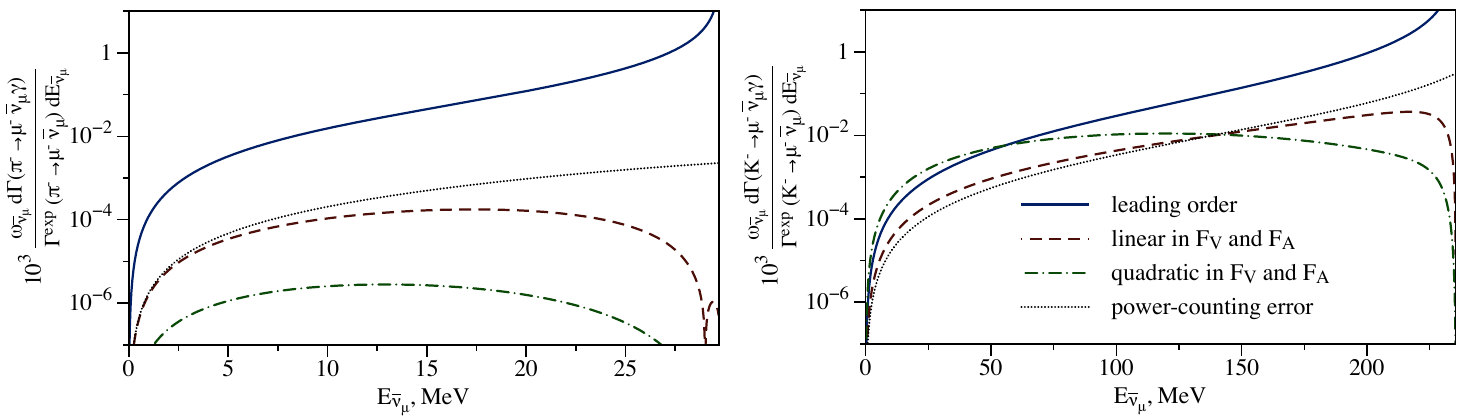}              
          \caption{Various contributions to the continuous (anti)neutrino energy spectra from the decay of light mesons normalized by $ \Gamma^{\mathrm{exp}}_{\pi^-,K^- \to \mu^- \bar{\nu}_\mu}/ \omega_{\bar{\nu}_\mu}$ are presented. Pion decay is presented on the left plot, kaon decay is shown on the right plot. Each graph compares the leading order (anti)neutrino energy spectrum $\mathrm{d} \Gamma \left( \pi^-,K^- \to \mu^- \bar{\nu}_\mu \gamma \right)/\mathrm{d} E_{\bar{\nu}_\mu}$ (blue solid line), the power-counting error estimate (black dotted line), and the strong dynamics contributions $\mathrm{d} | \Gamma_{F} \left( \pi^-,K^- \to \mu^- \bar{\nu}_\mu \gamma \right)| /\mathrm{d} E_{\bar{\nu}_\mu}$ (red dashed line) and $\mathrm{d} \Gamma_{F^2} \left( \pi^-,K^- \to \mu^- \bar{\nu}_\mu \gamma \right)/\mathrm{d} E_{\bar{\nu}_\mu}$ (green dashed-dotted line). Note: the power-counting error estimate is performed multiplying the contribution from non-eikonal terms of Eq.~(\ref{eq:leading_radiation}), i.e., beyond the first one, with $\frac{m_{\pi,K}^2}{16 \pi^2 f^2_{\pi,K}}$. \label{fig:pion_kaon_decay}}
\end{figure}
The decay spectrum is peaked approaching the kinematic endpoint of the process without radiation and behaves as $1/\left( \omega_{\bar{\nu}_\mu} - E_{\bar{\nu}_\mu} \right)$ near the endpoint, see Fig.~\ref{fig:pion_kaon_decay}. Integrating the (anti)neutrino energy spectrum up to $E_{\bar{\nu}_\mu} = \frac{1}{2} \left( m_\pi - \frac{m^2_\mu}{m_\pi - 2 \varepsilon}\right)$ regulates the dependence on the artificial cutoff parameter $\varepsilon$, c.f. Eq.~(\ref{eq:tree_level_soft_and_virtual_pion_decay_width_in_terms_of_the_experiment}), on the level of the decay width as well as in possible convolutions.\footnote{The integration over the $\epsilon$-suppressed region $ \omega_{\bar{\nu}_\mu} - \varepsilon \le E_{\bar{\nu}_\mu} \le \frac{1}{2} \left( m_\pi - \frac{m^2_\mu}{m_\pi - 2 \varepsilon}\right)$ contributes to the observables as 
\begin{equation}
\frac{\mathrm{d} \Gamma^\varepsilon \left( \pi^- \to \mu^- \bar{\nu}_\mu \gamma \right) }{\Gamma^{\mathrm{exp}}_{\pi^- \to \mu^- \bar{\nu}_\mu \left( \gamma \right)} \mathrm{d} E_{\bar{\nu}_\mu}}  = \frac{\alpha}{\pi} \left( 1+ \frac{m_\pi^2 + m_\mu^2}{m_\pi^2 - m_\mu^2} \ln \frac{m_\mu}{m_\pi} \right) \ln \frac{m^2_\mu}{m^2_\pi} \delta \left(E_{\bar{\nu}_\mu} -\left( \omega_{\bar{\nu}_\mu} -   \left( 1 - \frac{\omega_{\bar{\nu}_\mu}}{m_\pi} \right) \varepsilon \right) \right).
\end{equation}}
Outside the end-point region, the (anti)neutrino energy spectrum is suppressed by more than three orders of magnitude compared to the intensity of the monochromatic line. In the case of pion decay, the ChPT power counting uncertainties are slightly larger than the contribution of the linear in form factors $F_A$ and $F_V$. In the case of kaon decay, they have a similar size. For the kaon decay, the quadratic contribution from strong dynamics to the (anti)neutrino spectra exceeds the linear contribution, which indicates missing ingredients for the consistent series expansion as well as on the pure convergence of the $\mathrm{SU}\left(3\right)$ perturbative series.

As an illustrative example, we compare the flux-averaged cross sections on $^{40} \mathrm{Ar}$ nucleus
\begin{align}
\sigma^{^{40}\mathrm{Ar}}_{\bar{\nu}_\mu} &= \frac{\int \sigma^{^{40}\mathrm{Ar}}_{{\bar{\nu}_\mu}} \left( E_{\bar{\nu}_\mu} \right) \frac{\mathrm{d} \Gamma \left( \pi^- \to \mu^- \bar{\nu}_\mu \left( \gamma \right) \right) }{\mathrm{d} E_{\bar{\nu}_\mu}} \mathrm{d} E_{\bar{\nu}_\mu}}{\int \frac{\mathrm{d} \Gamma \left( \pi^- \to \mu^- \bar{\nu}_\mu \left( \gamma \right) \right) }{\mathrm{d} E_{\bar{\nu}_\mu}} \mathrm{d} E_{\bar{\nu}_\mu}}, \label{eq:flux_average}
\end{align}
accounting for the radiation of one photon in the (anti)neutrino production from the decay of the pion at rest $\sigma_{\bar{\nu}_\mu}^{^{40} \mathrm{Ar}}$ to the integration over the monochromatic spectrum $\sigma^{^{40} \mathrm{Ar}}_\mathrm{\bar{\nu}_\mu,LO}$~\cite{Payne:2019wvy,Yang:2019pbx,VanDessel:2020epd,Hoferichter:2020osn,Tomalak:2020zfh} with the same normalization in both the cases:
\begin{align}
\sigma^{^{40}\mathrm{Ar}}_{\bar{\nu}_\mu} &= \left(15.1867\pm0.25\right) \times 10^{-40}~\mathrm{cm}^2, \\
\sigma^{^{40}\mathrm{Ar}}_{\bar{\nu}_\mu,\mathrm{LO}} &= \left(15.1875\pm0.25\right) \times 10^{-40}~\mathrm{cm}^2.
\end{align}
The resulting relative effect of radiative corrections is around $6\times 10^{-5}$. The continuous component of the neutrino energy spectrum from the decay of light mesons can be safely neglected as soon as the flux is normalized to the meson decay width, accounting for all radiative events.

The effects of radiative corrections on accelerator (anti)neutrinos produced from pions decaying in flight are also negligibly small. As a toy example, let us consider a monochromatic beam of pions with energy $4.7~\mathrm{GeV}$. In the forward direction, such pions decay into (anti)neutrinos of $2~\mathrm{GeV}$ energy. In this setup, the radiative (anti)neutrino energy spectrum lowers the mean (anti)neutrino energy by a negligible $3.2\times10^{-5}$ relative correction, which is approximately independent of the off-axis angle on the length scales of modern detectors in oscillation neutrino experiments.

\section{Muon decay}
\label{sec3}

Pions and kaons decay into muons, which become a source of (anti)neutrinos of both electron and muon flavors. In this Section, we calculate (anti)neutrino energy spectra from the radiative decay of the muon $\mu^- \to e^- \bar{\nu}_e \nu_\mu (\gamma)$ and $\mu^+ \to e^+ {\nu}_e \bar{\nu}_\mu (\gamma)$ in the muon rest frame accounting for all electron-mass effects. Accounting for the isotropic invariance, neutrino energy spectra in the muon rest frame do not depend on the polarization of the decaying muon.

\subsection{Muon decay at tree level}
\label{sec3:mu_tree}

Muon decay $\mu^- \left( p_\mu \right) \to e^- \left( k_e \right) \bar{\nu}_e \left( k_{\bar{\nu}_e} \right) \nu_\mu \left( k_{\nu_\mu} \right)$  at leading order is governed by the low-energy effective four-fermion interaction with a scale-independent Fermi coupling constant $\mathrm{G}_\mathrm{F}$~\cite{Fermi:1934hr,Feynman:1958ty,Arason:1991ic,Antonelli:1980zt,Hill:2019xqk}
\begin{align}
    {\cal L}_{\rm eff} = - 2 \sqrt{2} \mathrm{G}_\mathrm{F} 
  \bar{\nu}_{\mu}\gamma^\mu \mathrm{P}_\mathrm{L} \nu_{e}
  \, \bar{e} \gamma_\mu \mathrm{P}_\mathrm{L}  \mu + \mathrm{h.c.}.
\end{align}
Performing the phase-space integration, we obtain the double-differential distribution w.r.t. the electron energy $E_e$ and (anti)neutrino energy $E_{\nu_\mu, {\bar{\nu}_e}}$, in terms of the squared matrix element at leading order $|\mathrm{T}_\mathrm{LO}|^2 =  64 \mathrm{G}_\mathrm{F}^2  p_\mu \cdot k_{\bar{\nu}_e} k_e \cdot k_{\nu_\mu}$:
\begin{align}
\frac{\mathrm{d} \Gamma_\mathrm{LO}\left( \mu^- \to e^- \bar{\nu}_e \nu_\mu \right)}{\mathrm{d} E_{\nu_\mu, {\bar{\nu}_e}} \mathrm{d} E_e} = \frac{|\mathrm{T}_\mathrm{LO}|^2 }{2^6 \pi^3 m_\mu}.
\end{align}
Integration of this distribution over the kinematically allowed range of electron energies
\begin{align}
\frac{q^4 + m^2_e m^2_\mu}{2 m_\mu q^2} \le E_e \le \frac{m^2_\mu + m^2_e}{2 m_\mu}, \label{eq:electron_energy_range}
\end{align}
results in the following (anti)neutrino energy spectra in the muon rest frame for electron and muon flavors~\cite{Okun:1982ap}:
\begin{align}
\frac{\mathrm{d} \Gamma_\mathrm{LO} \left( \mu^- \to e^- \bar{\nu}_e \nu_\mu \right)}{\mathrm{d} E_{\bar{\nu}_e}} &=  \frac{\mathrm{G}_\mathrm{F}^2 }{2 \pi^3} \frac{\left( q^2-m_e^2 \right)^2}{q^2}  E_{\bar{\nu}_e}^2, \label{eq:tree_level_muon_decay_width1} \\
\frac{\mathrm{d} \Gamma_\mathrm{LO} \left( \mu^- \to e^- \bar{\nu}_e \nu_\mu \right)}{\mathrm{d} E_{\nu_\mu}} &=  \frac{\mathrm{G}_\mathrm{F}^2 }{\pi^3}  \left( \frac{q^2-m_e^2}{q^2} \right)^2 \left( \frac{q^2 - m^2_e}{12} + \frac{q^2 +2 m^2_e}{6 q^2} m_\mu \left(m_\mu -  E_{\nu_\mu} \right) \right) E_{\nu_\mu}^2, \label{eq:tree_level_muon_decay_width2}
\end{align}
with the momentum transfer from the muon to the corresponding (anti)neutrino $q^2 =  m^2_\mu - 2 m_\mu E_{\nu_\mu, {\bar{\nu}_e}}$. The (anti)neutrino energy range is given by $ 0 \le E_{\nu_\mu, {\bar{\nu}_e}} \le \omega_{\nu} = \frac{m_\mu^2-m_e^2}{2m_\mu}$. Integrating over the (anti)neutrino energy, we obtain the well-known muon decay width at leading order
\begin{equation}
 \Gamma_\mathrm{LO} \left( \mu^- \to e^- \bar{\nu}_e \nu_\mu \right) = \frac{\mathrm{G}_\mathrm{F}^2 m^5_\mu}{192 \pi^3}\left( 1 - 8 r^2 - 24 r^4 \ln r + 8 r^6 - r^8  \right),
\end{equation}
with $r= \frac{m_e}{m_\mu}$.

At leading order, the radiative muon decay $\mu^- \left( p_\mu \right) \to e^- \left( k_e \right) \bar{\nu}_e \left( k_{\bar{\nu}_e}\right) {\nu}_\mu \left( k_{{\nu}_\mu}\right) \gamma \left( k_\gamma \right)$ is described by the Bremsstrahlung contribution $\mathrm{T}^{1\gamma}$~\cite{Kinoshita:1957zz}:
\begin{align}
\mathrm{T}^{1\gamma} = - 2 \sqrt{2} \mathrm{G}_\mathrm{F} i e ~\bar{\nu}_{\mu}\gamma^\mu \mathrm{P}_\mathrm{L} \nu_{e}
  \, \left(  \left( \frac{ {k}^\nu_e }{  k_e \cdot k_\gamma } - \frac{ {p}^\nu_\mu }{p_\mu \cdot k_\gamma } \right)  \bar{e} \gamma_\mu \mathrm{P}_\mathrm{L} \mu + \frac{1}{2} \bar{e}  \left( \frac{ \gamma^\nu \slash{k}_\gamma \gamma_\mu}{  k_e \cdot k_\gamma } + \frac{ \gamma_\mu \slash{k}_\gamma \gamma^\nu}{  p_\mu \cdot k_\gamma } \right)  \mathrm{P}_\mathrm{L} \mu \right) \varepsilon^\star_\nu.
\end{align}

In the following Sections, we include virtual corrections to the muon decay and perform the integration over the photon, electron, and (anti)neutrino phase space.

\subsection{Virtual corrections}
\label{sec3:mu_virtual}

To evaluate virtual contributions, it is convenient to express vertex corrections as a deviation of the charged-lepton current $\delta \mathrm{J}^\mathrm{L}_\nu$ from the tree-level expression $\mathrm{J}^\mathrm{L}_\nu = \bar{e} \left( k_e \right) \gamma_\nu \mathrm{P}_\mathrm{L} \mu \left( p_\mu \right)$ as
\begin{align}\label{eq:dJLR}
\delta \mathrm{J}^\mathrm{L}_\nu = e^2 \int \frac{\mathrm{d}^d L }{(2 \pi)^d} \frac{\bar{e} \left( k_e \right) \gamma^\lambda \left( \slash{k}_e - \slash{L} + m_e \right)  \gamma_\nu \mathrm{P}_\mathrm{L}  \left( \slash{p}_\mu - \slash{L} + m_\mu \right) \gamma^\rho \mu \left( p_\mu \right)}{\left((k_e - L)^2 - m_e^2 \right)\left( (p_\mu - L)^2 - m_\mu^2 \right)} \mathrm{\Pi}_{\lambda \rho} \left( L \right),
\end{align}
with the momentum-space photon propagator of Eq.~(\ref{eq:photon_propagator}). The corresponding field renormalization factors for external charged leptons are given above in Eq.~(\ref{eq:charged_lepton_Z_factor}). Neglecting Lorentz structures whose contractions with the (anti)neutrino current vanish at $m_\nu = 0$, the resulting correction to the charged lepton current is expressed as
\begin{align}
\left( \sqrt{Z_e Z_\mu} - 1 \right) \left(\mathrm{J}^\mathrm{L}\right)^\nu +  \left(\delta \mathrm{J}^\mathrm{L}\right)^\nu = \frac{\alpha}{2\pi} \bar{e} \left( k_e \right) \left( g_M \gamma^\nu - f_2 \frac{p_\mu^\nu+ r k_e^\nu}{2m_\mu} - g^5_M \gamma^\nu \gamma_5 - f^5_2 \frac{p_\mu^\nu - r k_e^\nu}{2m_\mu}   \gamma_5 \right) \mu \left( p_\mu \right), \label{eq:f1f2}
\end{align}
in terms of the form factors $g_M,~g^5_M,~f_2$, and $f^5_2$:
\begin{align}
g^{(5)}_M \left( \eta, r, \beta \right) &=- 1 + \frac{1}{\beta} \left( \frac{1}{2} \left(2 \beta -   \ln \frac{1+\beta}{1-\beta} \right)\ln \frac{2m_\mu}{\lambda}+ \frac{1}{2} \ln \frac{1+\beta}{1-\beta}  \ln \frac{1+\beta}{ \beta} -  \ln \frac{r \sqrt{1 - \beta} -  \sqrt{1 + \beta} }{  r \sqrt{1 + \beta}  - \sqrt{1 - \beta} } \frac{\ln r}{2} \right. \nonumber \\
&+ \left. \frac{3}{8}  \ln \frac{1+\beta}{1-\beta} + \frac{ \sqrt{1-\beta^2}}{8\eta}   \ln \frac{1+\beta}{1-\beta} + \frac{1}{4}  \ln \frac{1+\beta}{1-\beta}  \ln \frac{2 r  - \left(1 + r^2 \right)\sqrt{1 - \beta^2}}{ r^2 \left(1-\beta\right)} + \frac{\pi^2}{12}  \right. \nonumber \\
&+ \left. \frac{1}{2} \mathrm{Li}_2 \frac{1-\beta}{1+\beta}  - \frac{1}{2} \mathrm{Li}_2 \left( \frac{ \sqrt{1-\beta}}{\sqrt{1+\beta}} r \right) - \frac{1}{2} \mathrm{Li}_2 \left( \frac{\sqrt{1-\beta}}{\sqrt{1+\beta}} \frac{1}{r} \right)     - \frac{5}{16} \ln^2 \frac{1+\beta}{1-\beta} -  \frac{1}{4} \ln^2 r  \right) \nonumber \\
&+ \frac{\sqrt{1-\beta^2} }{8 \beta} \frac{  \left( 1 + \eta r \right)^2 \left( 1 - \eta \sqrt{1-\beta^2} \right) }{2r-\left( 1+ r^2 \right)\sqrt{1-\beta^2}}   \ln \frac{1+  \beta}{1-\beta}  + \frac{12 r - \left( 7 r^2 + 5 \right) \sqrt{1-\beta^2}  }{2r-\left( 1+ r^2 \right)\sqrt{1-\beta^2}} \frac{\ln r}{4} - \ln 2 r ,\label{DR_regularization} \\
f^{(5)}_2 \left( \eta, r, \beta \right) &= \frac{1}{2} \frac{\sqrt{1-\beta^2}}{\beta} \frac{1-\eta \sqrt{1-\beta^2}}{2r-\left( 1+ r^2 \right)\sqrt{1-\beta^2}} \ln \frac{1+\beta}{1-\beta} +  \frac{1 - \eta r}{1 + \eta r}  \frac{\sqrt{1-\beta^2}}{2r-\left( 1+ r^2 \right)\sqrt{1-\beta^2}} \ln r,
\end{align}
with the velocity of the electron in the muon rest frame $\beta$, $\eta = 1$ for the form factors $g_M,~f_2$ and $\eta = -1$ for the form factors $g^5_M,~f^5_2$.  As a cross-check of our calculation, we have verified the virtual one-loop QED contribution to the muon decay width~\cite{Berman:1958gx} in the limit of small electron mass, considering both neutrino and antineutrino energy spectra.

\subsection{Soft-photon Bremsstrahlung}
\label{sec3:mu_soft}

As for the radiative pion decay into soft photons, see Section~\ref{sec2:pi_soft}, the radiative decay width with photons below some small energy cutoff $k_\gamma \le \varepsilon$ (in the muon rest frame) factorizes in terms of the tree-level decay widths of Eqs.~(\ref{eq:tree_level_muon_decay_width1}) and~(\ref{eq:tree_level_muon_decay_width2}) as
\begin{align}
 \mathrm{d} \Gamma \left( \mu^- \to e^- \bar{\nu}_e \nu_\mu \gamma \left( k_\gamma \le \varepsilon \right) \right) =  \frac{\alpha}{\pi} \delta_s \left( \varepsilon \right)  \mathrm{d} \Gamma_\mathrm{LO} \left( \mu^- \to e^- \bar{\nu}_e \nu_\mu \right),\label{eq:soft_pion_decay_width}
\end{align}
with the universal soft-photon correction $\delta_s \left( \varepsilon \right)$ of Eq.~(\ref{eq:soft_result}), which depends on the electron velocity in the muon rest frame $\beta$. As a result, soft and virtual contributions multiply the tree-level spectra of Eqs.~(\ref{eq:tree_level_muon_decay_width1}) and~(\ref{eq:tree_level_muon_decay_width2}) as
\begin{align}
&\frac{\mathrm{d} \Gamma \left( \mu^- \to e^- \bar{\nu}_e \nu_\mu \right)+ \mathrm{d} \Gamma \left( \mu^- \to e^- \bar{\nu}_e \nu_\mu \gamma \left( k_\gamma \le \varepsilon \right) \right)}{\mathrm{d} \Gamma_\mathrm{LO} \left( \mu^- \to e^- \bar{\nu}_e \nu_\mu \right)}  = 1 + \frac{\alpha}{\pi} \left[ g_M + g^5_M + \delta_s \left( \varepsilon \right) \right. \nonumber \\
&\left. - \frac{r m^2_\mu}{4} \frac{\left( p_\mu - k_e \right)^2}{  p_\mu \cdot k_{\bar{\nu}_e} k_e \cdot k_{\nu_\mu} }  \left( g_M - g^5_M \right) + \left( \frac{r^2 m^2_\mu}{4} \frac{\left( p_\mu - k_e \right)^2}{p_\mu \cdot k_{\bar{\nu}_e} k_e \cdot k_{\nu_\mu}} -  \frac{ k_e \cdot k_{\bar{\nu}_e}    }{  p_\mu \cdot k_{\bar{\nu}_e} } \right) \left( \left( \frac{1+r}{2} \right)^2 f_2 + \left( \frac{1-r}{2} \right)^2 f^5_2 \right)  \right]. \label{eq:tree_level_soft_and_virtual_muon_decay_width}
\end{align}
To obtain the contributions to (anti)neutrino energy spectrum from virtual corrections and radiation of soft photons, we integrate these contributions over the electron energy in the kinematically allowed region of the process without radiation, c.f. Section~\ref{sec3:mu_tree} for technical steps.

\subsection{Contribution of hard photons}
\label{sec3:mu_hard}

To evaluate the radiation of photons above the small energy cutoff parameter $k_\gamma \ge \varepsilon$, we write down the squared matrix elements for the radiative muon decay amplitude $\mathrm{T}^{1\gamma}$, which is introduced in Section~\ref{sec3:mu_tree}, in terms of Lorentz invariants as
\begin{align}
\frac{|\mathrm{T}^{1\gamma}|^2 }{e^2 |\mathrm{T}_\mathrm{LO}|^2} &=  - \left( \frac{p_\mu}{p_\mu \cdot k_\gamma} - \frac{k_e}{k_e \cdot k_\gamma} \right)^2 +  \frac{p_\mu \cdot k_e}{p_\mu \cdot k_\gamma k_e \cdot k_\gamma } \left( \frac{k_{\nu_\mu} \cdot k_\gamma}{k_e \cdot k_{\nu_\mu}} -   \frac{k_{\bar{\nu}_e} \cdot k_\gamma }{p_\mu \cdot k_{\bar{\nu}_e}} \right) +  \frac{k_{\bar{\nu}_e} \cdot k_\gamma}{p_\mu \cdot k_{\bar{\nu}_e} p_\mu \cdot k_\gamma}  -  \frac{p_\mu \cdot k_{\nu_\mu}}{k_e \cdot k_{\nu_\mu} p_\mu \cdot k_\gamma} \nonumber \\
&+ \frac{1}{k_e \cdot k_\gamma} - \frac{1}{p_\mu \cdot k_\gamma} + \frac{k_{\nu_\mu} \cdot k_\gamma}{k_e \cdot k_{\nu_\mu} k_e \cdot k_\gamma}  + \frac{k_e \cdot k_{\bar{\nu}_e}}{p_\mu \cdot k_{\bar{\nu}_e} k_e \cdot k_\gamma} + \frac{k_{\bar{\nu}_e} \cdot k_\gamma }{ \left( p_\mu \cdot k_\gamma \right)^2 } \frac{m^2_\mu}{p_\mu \cdot k_{\bar{\nu}_e}} - \frac{ k_{\nu_\mu} \cdot k_\gamma }{ \left( k_e \cdot k_\gamma \right)^2 } \frac{m^2_e}{k_e \cdot k_{\nu_\mu}}. \label{eq:matrix_element_radiative_muon}
\end{align}
We follow the technique that was developed in~\cite{Ram:1967zza} and further exploited in~\cite{Sarantakos:1982bp,Passera:2000ug,Tomalak:2019ibg}. This technique was introduced for $2 \to 2$ scattering processes with radiation such as the elastic neutrino-electron scattering with one charged particle in the initial state and one charged particle in the final state. Performing the crossing of the initial neutral particle, we generalize the calculation to the decay process with radiation.

For the muon decay, we introduce the four-vector $l = p_\mu - k_{\nu_\mu} - k_{\bar{\nu}_e} = \left( l_0, \vec{f} \right)$ and the angle $\gamma$ between the photon momentum and the vector $\vec{f},~f=|\vec{f}|$. The energy-momentum conservation implies that
\begin{align}
l^2 - m^2_e = 2 k_\gamma \left( l_0 - f \cos \gamma \right).
\end{align}
Performing the integration over the electron momenta and angular variables of the (anti)neutrino of interest, we obtain the (anti)neutrino energy distribution
\begin{align}
\frac{\mathrm{d} \Gamma^{1\gamma}}{\mathrm{d} {E}_{\nu_\mu} \mathrm{d} {E}_{\bar{\nu}_e}}= \int \frac{|\mathrm{T}^{1\gamma}|^2}{2^{11} \pi^6 m_\mu} \frac{l^2 - m_e^2}{\left( l_0 - f \cos \gamma \right)^2}  \  f \mathrm{d} f \mathrm{d} \Omega_\gamma.\label{eq:radiative_muon_decay_width}
\end{align}
With our choice of the integration order, the integrand in Eq.~(\ref{eq:radiative_muon_decay_width}) does not depend on the photon azimuthal angle resulting in one trivial integration. Conveniently, we split the integration into two parts. There are no restrictions on the photon phase space in the region $\mathrm{I}$: $l^2 - m^2_e \ge 2 \varepsilon \left( l_0 - f \cos \gamma \right)$. For the muon neutrino energy spectrum, the range of kinematic variables in this region is given by
\begin{align}
0 \le E_{\nu_\mu}  \le \frac{m_\mu}{2} - \frac{1}{2} \frac{m^2_e}{m_\mu - 2 \varepsilon},  \\
0 \le  E_{\bar{\nu}_e} \le  \frac{m_\mu}{2} - \frac{1}{2} \frac{m^2_e}{m_\mu - 2 E_{\nu_\mu} - 2 \varepsilon},  \\
|  E_{\nu_\mu} - E_{\bar{\nu}_e} | \le f \le \mathrm{min} \left( E_{\nu_\mu} + E_{\bar{\nu}_e},~ \sqrt{ \left( l_0 - \varepsilon \right)^2- m^2_e} - \varepsilon \right).
\end{align}
To obtain the electron antineutrino energy spectrum, we have to substitute $E_{\nu_\mu} \leftrightarrow E_{\bar{\nu}_e}$ in all kinematic ranges. In the region $\mathrm{II}$: $l^2 - m^2_e \le 2 \varepsilon \left( l_0 - f \cos \gamma \right)$, the angle between the photon momentum and the vector $\vec{f}$ is constrained as
\begin{align}
\cos \gamma  \ge \frac{1}{f} \left( l_0 - \frac{l^2-m_e^2}{2\varepsilon} \right).
\end{align}
The phase space of the region $\mathrm{II}$ is close to the kinematics of the process without Bremsstrahlung. Only the first term from Eq.~(\ref{eq:matrix_element_radiative_muon}) contributes in this region. Moreover, only this factorizable term generates the dependence on the photon energy cutoff $\varepsilon$ from region $\mathrm{I}$, which cancels the contribution of low-energy photons with $k_\gamma \le \varepsilon$. Considering the nonfactorizable radiation terms, we safely set $\varepsilon = 0$ in the beginning of the calculation. 

We have reproduced the known contribution of photons with energy $k_\gamma \ge \varepsilon$ to the muon decay width~\cite{Kinoshita:1959uwa} in the limit of small electron mass. As another cross-check of our calculation, we have verified the Bremstrahlung $\mathrm{O} \left( \alpha \right)$ contribution to the muon decay width~\cite{Berman:1958gx} in the limit of small electron mass, considering both neutrino and antineutrino energy spectra.

\subsection{Decay spectra and integrated cross sections}
\label{sec3:mu_results}

Combining all pieces together, we have reproduced $\mathrm{O} \left( \alpha \right)$ contribution to the muon lifetime both for small and finite electron masses~\cite{Berman:1958gx,Kinoshita:1958ru,Nir:1989rm,vanRitbergen:1999fi,Pak:2008qt,Sirlin:2012mh}:
\begin{equation}
  \Gamma \left( \mu^- \to e^- \bar{\nu}_e \nu_\mu \left( \gamma \right) \right) = \Gamma_\mathrm{LO} \left( \mu^- \to e^- \bar{\nu}_e \nu_\mu \right)  + \frac{\alpha}{\pi} \left(  \frac{25}{8} - \frac{\pi^2}{2} - \left( 34 + 24 \ln r \right) r^2  + 16 \pi^2 r^3 + ... \right) \frac{\mathrm{G}_\mathrm{F}^2 m^5_\mu}{192 \pi^3}.
\end{equation}
Thus, we consider the effect of radiative corrections on (anti)neutrino energy spectra and flux-averaged cross sections. In the limit of vanishing electron mass, we obtain radiative electron antineutrino and muon neutrino energy spectra in agreement with QCD corrections to the lepton energy spectra in decays of heavy quarks~\cite{Jezabek:1988ja}
\begin{align}
\frac{\mathrm{d} \Gamma \left( \mu^- \to e^- \bar{\nu}_e \nu_\mu \left( \gamma \right) \right)}{\mathrm{d} \Gamma_\mathrm{LO} \left( \mu^- \to e^- \bar{\nu}_e \nu_\mu \right)} &= - \frac{\alpha}{\pi}  \left( \mathrm{Li}_2 \frac{2 E_{\bar{\nu}_e}}{m_\mu}  + \frac{1}{2} \ln^2 \left( 1 - \frac{2 E_{\bar{\nu}_e}}{m_\mu} \right) + \frac{\pi^2}{3} - \frac{19}{24} + \frac{5}{24} \frac{m_\mu}{E_{\bar{\nu}_e}} \right. \nonumber \\
& \left.+ \left( \frac{2}{3} + \frac{1}{3} \frac{m_\mu}{E_{\bar{\nu}_e}} + \frac{5}{48} \frac{m_\mu^2}{E^2_{\bar{\nu}_e}} \right) \ln \left( 1 - \frac{2 E_{\bar{\nu}_e}}{m_\mu} \right) \right), \label{eq:antineutrino} \\
\frac{\mathrm{d} \Gamma \left( \mu^- \to e^- \bar{\nu}_e \nu_\mu \left( \gamma \right) \right) }{\mathrm{d}  \Gamma_\mathrm{LO} \left( \mu^- \to e^- \bar{\nu}_e \nu_\mu \right) } &= - \frac{\alpha}{\pi}  \left( \mathrm{Li}_2 \frac{2 E_{\nu_\mu}}{m_\mu}  + \frac{1}{2} \ln^2 \left( 1 - \frac{2 E_{\nu_\mu}}{m_\mu} \right) + \frac{\pi^2}{3} + \frac{ \frac{43}{6} \frac{E_{\nu_\mu}}{m_\mu}  - \frac{51}{8} + \frac{41}{24} \frac{m_\mu}{E_{\nu_\mu}}}{3 - \frac{4 E_{\nu_\mu}}{m_\mu}} \right. \nonumber \\
& \left.- \frac{ \frac{8}{3} \frac{E_{\nu_\mu}}{m_\mu}  - \frac{7}{2} + \frac{3}{2} \frac{m_\mu}{E_{\nu_\mu}} - \frac{41}{48} \frac{m_\mu^2}{E^2_{\nu_\mu}}}{3 - \frac{4 E_{\nu_\mu}}{m_\mu}} \ln \left( 1 - \frac{2 E_{\nu_\mu}}{m_\mu} \right) \right). \label{eq:neutrino} 
\end{align}

In the following Fig.~\ref{fig:muon_decay}, we compare the leading-order (anti)neutrino energy spectrum from the muon decay $\mathrm{d} \Gamma_\mathrm{LO} \left( \mu^- \to e^- \bar{\nu}_e \nu_\mu \right)/\mathrm{d} E_{\nu_\mu, {\bar{\nu}_e}}$ of Eqs.~(\ref{eq:tree_level_muon_decay_width1}, \ref{eq:tree_level_muon_decay_width2}) to the contribution of radiative corrections at $\mathrm{O} \left( \alpha \right)$ level from Eqs.~(\ref{eq:antineutrino}) and (\ref{eq:neutrino}) in the limit of a massless electron as well as to corrections in the electron mass at $\mathrm{O} \left( \alpha \right)$ level.
\begin{figure}[t]
          \centering
          \includegraphics[height=0.289\textwidth]{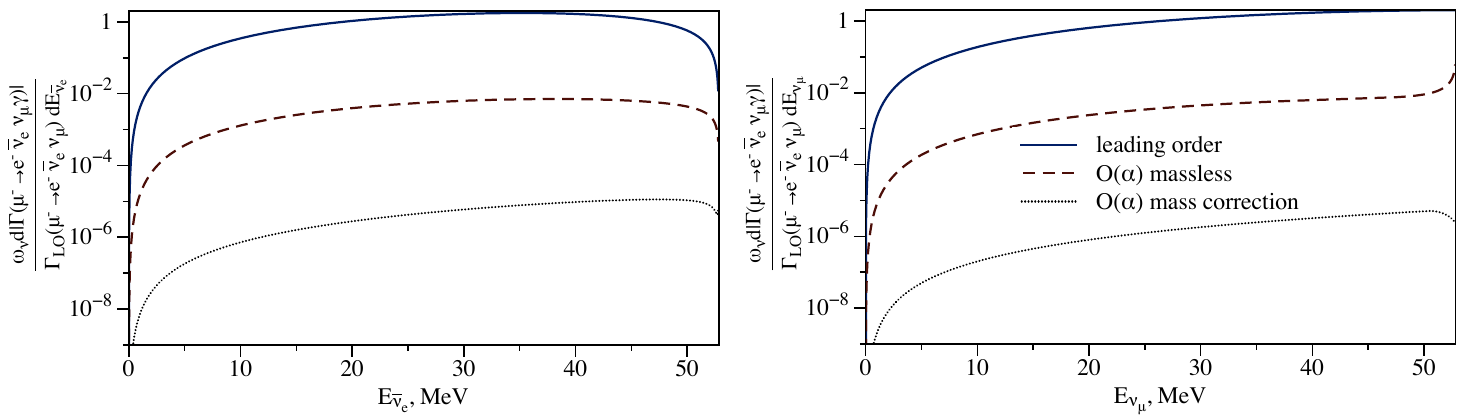}              
          \caption{(Anti)neutrino energy spectra from the decay of the negative muon $\mu^- \to e^-  \bar{\nu}_e \nu_\mu \left( \gamma \right)$ is shown normalized by $\Gamma_\mathrm{LO} \left( \mu^- \to e^- \bar{\nu}_e \nu_\mu \right)/ \omega_{\nu}$. The electron antineutrino energy spectrum is presented on the left plot, and the muon neutrino energy spectrum is shown on the right plot. The leading-order result $\Gamma_\mathrm{LO} \left( \mu^- \to e^- \bar{\nu}_e \nu_\mu \right)$ is shown by the blue solid line. It is compared to the $\mathrm{O} \left( \alpha \right)$ contribution of Eqs.~(\ref{eq:antineutrino}) and~(\ref{eq:neutrino}) for a massless electron (the red dashed line) and to the electron-mass corrections at $\mathrm{O} \left( \alpha \right)$ order (the black dotted line).\label{fig:muon_decay}}
\end{figure}
As expected from the QED power counting, radiative corrections contribute around $3$-$4$ permille in the dominant energy region of the (anti)neutrino flux. (Anti)neutrino energy spectra are infrared-safe observables, which are free from $\ln \frac{m_e}{m_\mu}$ singularities according to the Kinoshita-Lee-Naunberg theorem~\cite{Bloch:1937pw,Yennie:1961ad,Kinoshita:1962ur,Lee:1964is,Greub:1993kg,Sirlin:2011wg}. Electron-mass corrections can be safely neglected because they are numerically of the next order in $\alpha$ expansion or even below the next order.

As an illustrative example, we compare the flux-averaged cross sections on $^{40} \mathrm{Ar}$ nucleus, with the averaging according to Eq.~(\ref{eq:flux_average}), accounting for the radiation of one photon in the (anti)neutrino production from the decay of the muon at rest $\sigma^{^{40} \mathrm{Ar}}$ to cross sections averaged over the leading-order flux $\sigma^{^{40} \mathrm{Ar}}_\mathrm{LO}$~\cite{Payne:2019wvy,Yang:2019pbx,VanDessel:2020epd,Hoferichter:2020osn,Tomalak:2020zfh}:
\begin{align}
\sigma_{\bar{\nu}_e}^{^{40}\mathrm{Ar}} &= \left(17.484\pm0.43\right) \times 10^{-40}~\mathrm{cm}^2, \qquad &\sigma_{\nu_\mu}^{^{40}\mathrm{Ar}} = \left(22.448\pm0.66 \right) \times 10^{-40}~\mathrm{cm}^2, \\
\sigma_{\bar{\nu}_e,\mathrm{LO}}^{^{40}\mathrm{Ar}} &= \left(17.490\pm0.43\right) \times 10^{-40}~\mathrm{cm}^2, \qquad &\sigma^{^{40}\mathrm{Ar}}_{\nu_\mu,\mathrm{LO}} = \left(22.454\pm0.66\right) \times 10^{-40}~\mathrm{cm}^2.
\end{align}
The resulting relative effect of radiative corrections is around 3 permille both for the electron antineutrino and for muon neutrino. Considering the same normalization for the leading-order and radiatively corrected spectrum, the relative effect changes sign and reduces to 1 permille. Consequently, to achieve control over the artificial neutrino flux at permille level, it is necessary to first account for the radiative corrections from the muon decay.

\section{Conclusions and Outlook}
\label{sec4}

The goal of this paper is to enable sub-percent control over (anti)neutrino energy distributions by improving the precision of artificial neutrino fluxes at low and high energies. Thus, we performed a study of radiative corrections to the main (anti)neutrino production channels, i.e., pion, kaon, and muon decays. Here we presented analytical expressions for (anti)neutrino energy spectra and compared leading-order fluxes to $\mathrm{O} \left(\alpha \right)$ contributions. Because experiments do not detect photons at the neutrino production sources, we included the radiation of all hard photons allowed by the kinematics of the decay. As proof-of-concept, we demonstrated the effect of radiative corrections on the scattering cross sections on $^{40} \mathrm{Ar}$ nucleus. All our results were obtained in the rest frame of the decaying particle; however, the results could easily be reproduced for an arbitrary rest frame performing the corresponding Lorentz boost.

Radiative corrections in decays of charged pseudoscalar mesons introduce a continuous and divergent near the (anti)neutrino endpoint component, on top of the monochromatic tree-level spectra. However, the relative intensity of the radiative tail is well below the strength of the monochromatic line. The flux-averaged coherent elastic neutrino-nucleus scattering cross section on $^{40} \mathrm{Ar}$ changes by the negligibly small amount of $6\times10^{-5}$ compared to the leading-order prediction, when the same normalization is used both for radiatively-corrected and tree-level predictions. Pion and kaon structure-dependent contributions are suppressed by at least two orders of magnitude compared to the leading point-like $\mathrm{O} \left(\alpha\right)$ QED effects and are irrelevant for the (anti)neutrino production mechanism.

Radiative corrections to (anti)neutrino spectra from the muon decay are around $3$-$4$ permille in the dominant region of the neutrino flux. Such effects introduce the similar size of the distortion for the flux-averaged cross sections on the nuclear target. The relative correction diverges near the endpoint of the (anti)neutrino energy, where the leading-order contribution vanishes, resulting into finite physical observables at the fixed order of the perturbation theory. 

(Anti)neutrino energy spectra are infrared-safe observables that are free from collinear logarithms of the $\ln \frac{m_e}{m_\mu}$ type. QED radiative corrections to such observables are well estimated by QED power counting and have a size of $\frac{\alpha}{\pi} \sim 2$ permille up to the factor of the natural size. Our results for (anti)neutrino spectra from the muon decay confirm this simple estimate. $\mathrm{O} \left(\alpha\right)$ electron-mass effects are a few orders of magnitude below the leading contribution and can be neglected for applications at the neutrino scattering facilities.

\section*{Acknowledgments}
O.T. thanks Ryan Plestid as well as Pedro Machado and Vishvas Pandey for numerous discussions on~\cite{Tomalak:2020zfh}, which has motivated the main part of this work, Vincenzo Cirigliano and Emanuele Mereghetti for useful advices regarding literature and presentation, Emanuele Mereghetti for comments on the text of this manuscript, Richard Hill, Matthias Heller, and Marc Vanderhaeghen for technical discussions while working on other projects, and Kevin McFarland for suggestions regarding applications to accelerator neutrino sources.  O. T. acknowledges the Fermilab theory group for warm hospitality and support. O. T. acknowledges the theory group of Institute for Nuclear Physics at Johannes Gutenberg-Universit\"at Mainz for warm hospitality and support. This work is supported by the US Department of Energy through the Los Alamos National Laboratory. Los Alamos National Laboratory is operated by Triad National Security, LLC, for the National Nuclear Security Administration of U.S. Department of Energy (Contract No. 89233218CNA000001). This research is funded by LANL’s Laboratory Directed Research and Development (LDRD/PRD) program under project number 20210968PRD4. The work of O.T. is partially supported by the Visiting Scholars Award Program of the Universities Research Association. FeynCalc~\cite{Mertig:1990an,Shtabovenko:2016sxi}, LoopTools~\cite{Hahn:1998yk}, Mathematica~\cite{Mathematica} and DataGraph were extremely useful in this work.

\bibliography{decays}{}

\end{document}